\documentclass[12pt,tightenlines,eqsecnum,floats,showpacs,nofootinbib,amsmath,amssymb,aps,prd,longbibliography]{revtex4-2}

\usepackage{tikz}
\usepackage{graphicx,verbatim}
\usepackage{amsmath}
\usepackage{amsfonts}
\usepackage{amssymb}
\usepackage{psfrag}
\usepackage{accents}
\usepackage{mathrsfs}
\usepackage{enumitem}
\usepackage{hyperref}
\usepackage{caption}

\newcommand*{\scrip}{\ensuremath{\mathscr{I}^{+}}} 
\newcommand*{\scrim}{\ensuremath{\mathscr{I}^{-}}} 

\usepackage[colorinlistoftodos]{todonotes}
\usepackage{epstopdf}
\usepackage{mathrsfs}

\def\be{\begin{equation}}
\def\ee{\end{equation}}
\def\ba{\begin{eqnarray}}
\def\ea{\end{eqnarray}}

\newcommand{\pb}[1]{\hbox{\lower0.5ex\hbox{${}_{\leftarrow}$}}\kern-1.9ex{#1}}

\def\={\,\hat{=}\,}
\def\f{\frac}
\def\B{\mathfrak{B}} 
\def\EH{\mathfrak{eh}}
\def\QLH{\mathfrak{h}}
\def\DHS{\mathfrak{DH}}
\def\IHS{\mathfrak{iH}}
\def\I{\mathtt{I}}
\def\S{\mathtt{S}}
\def\Q{\mathtt{Q}}

\usepackage{pgf}
\makeatletter
\newcommand*{\pgfunderleftarrow}{%
  \@ifstar
    {\let\ifpgf@depth\iftrue\mathpalette\@pgfunderleftarrow}
    {\let\ifpgf@depth\iffalse\mathpalette\@pgfunderleftarrow}%
}
\newcommand*{\@pgfunderleftarrow}[2]{%
  #2%
  \edef\pgf@math@fam{\the\fam}%
  \pgfpicture
    \pgfsetbaseline{0pt}
    \pgf@relevantforpicturesizefalse      
    \pgfsetroundcap                       
    \pgfsetarrowsend{to}
    \pgfutil@tempdima=0.28pt%
    \advance\pgfutil@tempdima by.8\pgflinewidth%
    \pgfutil@tempdima-4\pgfutil@tempdima
    \sbox\pgfutil@tempboxa{$\m@th\fam\pgf@math@fam#1#2$}%
    \advance\pgfutil@tempdima-\dp\pgfutil@tempboxa
    \pgfutil@tempdimb\wd\pgfutil@tempboxa
    \pgfpathmoveto{\pgfqpoint{0pt}{\pgfutil@tempdima}}%
    \pgfpathlineto{\pgfqpoint{-\pgfutil@tempdimb}{\pgfutil@tempdima}}%
    \pgfusepath{stroke}
    \ifpgf@depth
      \pgf@relevantforpicturesizetrue
      \pgfpathmoveto{\pgfqpoint{0pt}{-\pgfutil@tempdimb}}%
      \pgfusepath{use as bounding box}%
    \fi
  \endpgfpicture
}
\makeatother

\begin{document}

\title{Surprises in Gravitational Physics}

\author{Abhay Ashtekar}
\email{ashtekar.gravity@gmail.com}
\affiliation{Physics Department, The Pennsylvania State 
University,\\ University Park, PA 16802, USA}
\begin{abstract}
\medskip

\noindent 

General relativity continues to surprise us, especially in its strong-field regime. These surprises have repeatedly forced us not only to broaden familiar paradigms, but sometimes to shift them. I will present a few examples to illustrate the unexpected vistas that can suddenly open up, bringing a sense of wonder, fascination, and respect for the profound subtlety of the theory. Because this volume is addressed to non-experts, I will emphasize the context in which these features arose and explain the underlying reasoning in general terms, rather than give proofs with their attendant technical caveats.
\end{abstract}

\maketitle

\section{Introduction}
\label{s1}

It is a pleasure to contribute to this memorial volume commemorating Prof. Jayant Narlikar's life and work.  I was fortunate to be associated with him at several stages of our lives. I attended high school in Kolhapur, India where the Narlikar family was well-known because of their past association with the town. Already then I knew that Jayant had been a `wrangler' in the Mathematics Tripos at Cambridge and had gone on to carry out forefront research in cosmology there. Later, starting in the mid-1980s, I came to know him personally when he founded IUCAA in Pune, where I visited him many times. During these visits, I came to know how passionate he was about science outreach.  Therefore, I have tried to make this tribute sufficiently non-technical to be accessible to a broad audience, while retaining enough detail to engage students interested in fundamental physics. 

With this goal in mind, I will discuss a few among the many surprises that I encountered over the last 50 years or so while studying and investigating the nature of strong gravity. On quite a few occasions, I came across a sudden, unexpected panorama, as one often does while climbing a mountain. Each time I was elated---`lifted up', so to say---and filled with a sense of awe at the interwoven subtlety of general relativity as it presents us with novel phenomena in the strong-field regime. In Francis Bacon's words, these sudden manifestations bring out the \emph{strangeness in proportion} of Einstein's theory, a hallmark of profound beauty.

Given the page limit, I will confine myself to two topics: black holes and gravitational waves beyond perturbation theory. They illustrate the uncanny ability of Einstein's equations to embed physics into space-time geometry.

\section{Black Holes}
\label{s2}

Although the possibility of black holes was already discussed by Laplace in the late 1790s, an examination of his argument shows that it is flawed: His black holes can be black for all observers only if the velocity of light is an absolute constant, independent of the state of motion of observers, which is not possible within the Newtonian framework he used. One needs general relativity, a theory of gravity in which the speed of propagation of light is an invariant notion. Indeed, even after the advent of general relativity, it took several decades for researchers to grasp the idea that there may be black holes in our universe. Even though Schwarzschild discovered his solution already in 1916 \cite{Schwarzschild1916}, it was generally believed that it was singular at $r={2GM}/{c^2}$, and therefore could only represent the exterior gravitational field of a spherical star whose radius would vastly exceed this value. 

In fact, two decades later, in 1939, Einstein himself argued that black holes cannot form in Nature \cite{Einstein1939}. His analysis was self-consistent but the conclusion was flawed because he focused on a stationary system of spherically distributed particles and the assumption of stationarity ruled out gravitational collapse. However, only a few months later, Oppenheimer and Snyder \cite{Oppenheimer:1939ue} showed that a spherical star consisting of pressureless fluid would undergo collapse when evolved using Einstein's equations, leading to what is now recognized as a black hole. But because black holes seemed exotic, there was great resistance among physicists, and especially astronomers, to taking them seriously as physical entities that can exist in Nature. 

This status quo continued well into the late 1970s. Today, the situation is just the opposite: Now it is `obvious' that black holes are ubiquitous in the universe! While many factors contributed to this transformation, I think the seminal role was played by the LIGO-Virgo observations of gravitational waves emitted during binary black hole mergers.  Interestingly, the first event was observed in 2015 \cite{Abbott:2016apu}, exactly a hundred years after Einstein's discovery of general relativity. In this section, I will discuss three fundamental features of black holes that greatly surprised the scientific community and also led to some confusion!
 \vskip0.3cm
\textbf{1. The Astonishing Simplicity of Black Holes}
\vskip0.1cm
\noindent Stationary stars exhibit a rich variety of shapes. They can be spherical, elliptic and, generically, oblate spheroids. Since they admit a rich variety of shape and spin multipole moments, it came as a major surprise when it was shown 
 that the situation is drastically different for black holes: stationary black holes are fully characterized by their mass $M$ and spin $J$ (see \cite{Chruciel2012} for a review). 
\footnote{I will restrict myself to four-dimensional space-times and, for brevity of presentation, assume that the cosmological constant and charges (carried by gauge fields) are zero. Even then the uniqueness theorem is firmly established only for stationary, axisymmetric black holes. Beyond axisymmetry the result requires certain technical assumptions, on which there is an ongoing debate. In what follows, I will ignore this caveat.}
All higher multipole moments are specific functions of just these two numbers. What is more, the solution of Einstein's equations representing this two-parameter family of black holes is explicitly known: It is the Kerr family \cite{Kerr1963}. Einstein's equations `know' that these fascinating physical entities exist and, furthermore, inform us that their equilibrium states are completely determined by the values of $M$ and $J$! Note that a black hole is an extreme physical manifestation of geometry; the only input required for its full description is the space-time metric.  From this metric emerge diverse unforeseen phenomena: the Penrose process through which one can continually extract energy from rotating black holes until they become spherical; the surprising similarity between the laws they obey and the fundamental laws of thermodynamics; and the quantum mechanical black hole radiance discovered by Hawking. As S. Chandrasekhar, the preeminent astrophysicist of the 20th century, put it \cite{Chandrasekhar:1985kt}: \vskip0.1cm
\textsl{``The black holes of nature are the most perfect macroscopic objects there are in the universe: the only elements in their construction are our concepts of space and time. And since the general theory of relativity provides only a single family of solutions for their descriptions, they are the simplest objects as well." }

Interestingly, and rather ironically, the first reaction of the community to the black hole uniqueness theorems was to conclude that black holes do \emph{not} exist in Nature. I will illustrate the logic using the non-rotating Schwarzschild black hole which is spherically symmetric. Oppenheimer and Snyder had shown that gravitational collapse leads to the formation of a Schwarzschild black hole. But their analysis \emph{began} with a perfectly spherical star. However, in the real world, non-rotating stars are never \emph{exactly} spherically symmetric. Therefore, the reasoning went, even if a real, astrophysical non-rotating star were to collapse the resulting object would not be spherically symmetric either. But, by uniqueness, the only non-rotating stationary black hole is Schwarzschild  and therefore spherically symmetric. So a black hole cannot be the end result. The end result would therefore be an object that retains irregularities---i.e., has non-vanishing shape multipoles. For rotating black holes, the argument is similar. Rotating stars have a variety of multipole moments while the multipoles of a rotating black hole are all \emph{specific} functions of just two parameters, $M$ and $J$. Therefore, the end result of a gravitational collapse of such a generic star would not be a Kerr black hole!

It took some time for the general relativity community to convince astronomers that this reasoning is flawed. Specifically, Penrose forcefully argued that black hole uniqueness simply means that multipoles would be radiated away through neutrino, electromagnetic, and gravitational radiation during the collapse. For a non-rotating star, \emph{all} the shape multipoles (except the monopole) would be radiated away, while for a rotating star, the \emph{departures} of its mass and spin multipoles from those of the final Kerr black hole would be. However, this possibility was largely ignored, probably regarded as a bold speculation. It took a few decades for this `radical' view to become `mainstream'.  The process was facilitated by the discovery that black holes can serve as  engines for the most energetic phenomena seen in the universe and the realization that mechanisms that do not involve black holes cannot account for these extreme events. But, as I mentioned above, it was finally cemented by the LIGO-Virgo discoveries of black hole mergers.

\vskip0.3cm
\textbf{2. Black Hole Thermodynamics}
\vskip0.2cm
\noindent
\begin{minipage}{0.25\linewidth}
  \centering
  \includegraphics[width=\linewidth]{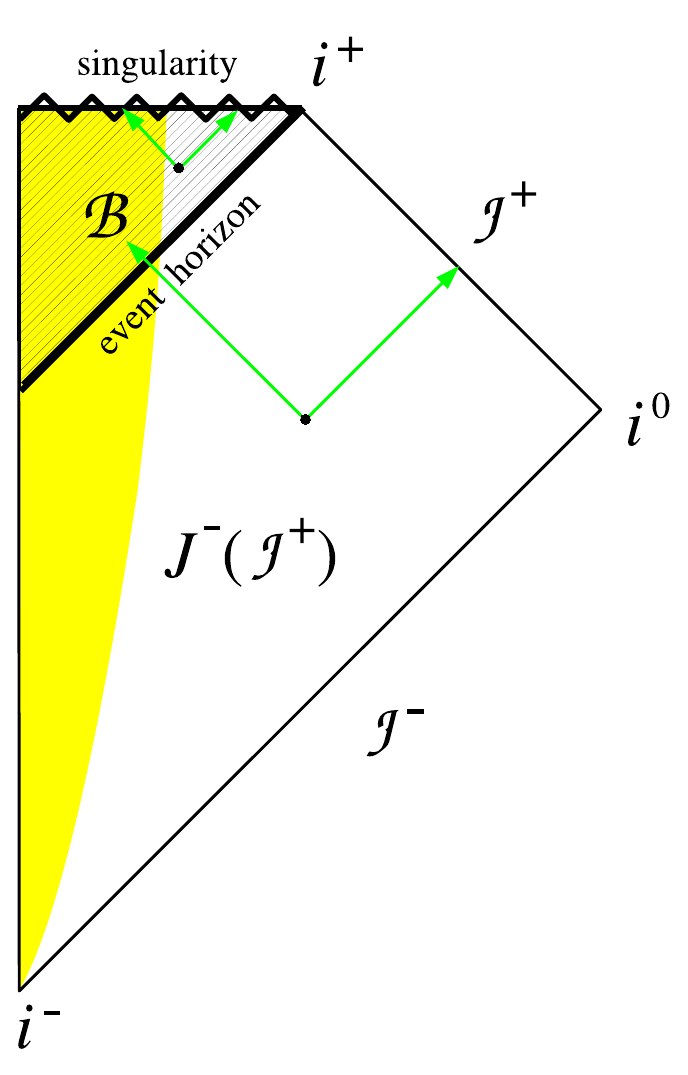}\vskip-0.7cm
   \footnotesize{Fig. 1} 

  \label{fig:1}
\end{minipage}
\hfill
\begin{minipage}{0.71\linewidth} \noindent The second conceptual surprise is associated with the form of the laws that black holes themselves obey. They have an astonishing similarity with the fundamental laws of thermodynamics, even though they result simply from the constraints on permissible geometries imposed by general relativity alone, \emph{without any reference to thermodynamics!} \vskip0.1cm

Let us begin with the basic notions involved. Figure 1 is the Penrose diagram of a space-time depicting a collapsing spherical star. The future null boundary of space-time is labeled $\scrip$. The asymptotic region (labeled $J^-({\scrip})$ in the figure) is the part of space-time from which future-directed causal signals can eventually reach $\scrip$, while the black hole region ($\mathcal{B}$) is the part from which they cannot, \emph{no matter how long one waits}. \end{minipage}
\vskip0.3cm

The three-dimensional null surface that separates the asymptotic and black hole regions is called the \emph{event horizon} and is taken to represent the black hole boundary. (See Fig. 1). The space-time metric in the source-free (unshaded) part is just the static Schwarzschild metric, and the portion of the event horizon that lies in this part represents a black hole boundary \emph{in equilibrium}. Kerr metrics represent eternal, stationary  black holes in space-times without any matter at all. Their entire black hole horizon is in equilibrium. 

Let us begin with these equilibrium configurations. Now, the event horizon of a generic Kerr black hole is not spherically symmetric; all of its shape multipoles are non-zero, bringing out the fact that the event horizon is quite distorted. Yet, the surface gravity $\kappa$ on these horizons is constant. This is in striking contrast with the surface gravity $g$ on the surface of the Earth which varies along the Earth's surface because the Earth is not exactly spherically symmetric. Thus, constancy of surface gravity $\kappa$ for black holes in equilibrium is an unforeseen property, not shared by other stationary astrophysical objects. 

Next, since for these black holes the entire space-time geometry is stationary, it follows that the area $A$ of a two-dimensional cross section $S$ of their event horizon does not change in time. Similarly, the angular velocity of the horizon, $\Omega$, spin $J$ and mass $M$ are also time independent. Thus, for black holes in equilibrium, $\kappa,\, A,\, \Omega, J$ and $M$ are just numbers associated with each equilibrium state. Let us now consider two equilibrium states of black holes that differ only infinitesimally (e.g., in the sense that their shape and  spin multipoles are very close to each other). Denote the infinitesimal differences in their mass, event horizon area and angular momentum by $(\delta M,\, \delta A, \,\delta J)$ . Then Einstein's  equations imply that there is a specific relation among these quantities: 
\be \label{1law1} \delta M = \f{\kappa}{8\pi G}\, \delta A + \Omega\, \delta J\,\, \equiv\,\, \f{\kappa}{8\pi G}\, \delta A + \hbox{\rm work required to change $J$}\,. \ee
These relations%
\footnote{This is a restriction to source-free space-times with stationary black holes of the law that Bardeen, Carter and Hawking obtained. They allowed stationary perfect fluids in space-times which complicates the analysis. For our purposes that complication is not necessary.}, 
discovered by Bardeen, Carter and Hawking \cite{BCH}, bear a striking similarity with the zeroth and the first laws of thermodynamics that also deal with systems in equilibrium: The temperature $T$ of a thermodynamic system in equilibrium is a constant and the energy, entropy, and volume of two nearby equilibrium states satisfy:
\be \label{1law} \delta E = T\, \delta S\, - p\, \delta V\,  \equiv T\, \delta S\, + \hbox{\rm work required to change $V$}\, . \ee 
Comparing the forms of (\ref{1law1}) and (\ref{1law}), one sees that $(M,\, A,\, J)$ of black holes in equilibrium are analogous to the extensive observables $(E,\,S,\, V)$ of thermodynamic systems in equilibrium and  $(\kappa,\, \Omega)$ are analogous to their intensive parameters $(T,\, p)$. Finally, following Penrose's ideas which appeared in print only later \cite{Penrose1974}, Hawking showed yet another striking consequence of Einstein's equations, now for dynamical black holes \cite{Hawking:1971vc}.%
\footnote{A historical account of this development was presented by  Klass Landsman at the July 2026 Oxford Conference \emph{Penrose-fest @ 95} and will appear in a book written jointly with Jeroen van Dongen, that is in preparation.}
Assuming cosmic censorship --the hypothesis that singularities of general relativity that can destroy predictivity are `clothed' behind horizons-- and assuming that sources satisfy an energy condition expected of physically reasonable classical matter, for any two cross-sections of the event horizon, one has:

\be \label{2law1} A(S_2)\, \ge\, A(S_1) \hskip0.5cm  \hbox{if $S_2$ lies to the causal future of $S_1$}\,. \ee
This relation is again strikingly similar to the second law of thermodynamics that asserts that entropy of a closed system cannot decrease in time. These similarities suggest one should associate with a black hole in equilibrium a temperature proportional to its surface gravity $\kappa$, and entropy proportional to the area $A$ of its event horizon 
\footnote{It is interesting to note that, a quarter century after the initial papers \cite{Penrose1974,Hawking:1971vc}, it was realized that the smoothness of event horizons, used crucially in proofs of the second law, \emph{fails} in the case of interest: dynamical black holes, such as those that merge \cite{Giulini1998,Krolak1999}! A version of the theorem that does hold generically was proved only in 2001 by Chrusciel, Delay, Galloway and Howard \cite{CDGH}, but it involves a measure-theoretic---rather than Riemannian-geometric---notion of area.}.
Around the same time as this discovery, Wheeler and Bekenstein used thought experiments (involving quasi-stationary processes from one equilibrium state to a nearby one) to argue that the second law of thermodynamics could be adequately generalized to accommodate black holes, provided one assigns an entropy to them, again proportional to the horizon area $A$ \cite{Bekenstein:1973ur,Bekenstein:1974ax}. (Hawking's subsequent discovery \cite{Hawking:1974sw} of quantum radiation by black holes fixed the proportionality factors, leading to $T = \kappa\hbar/2\pi$ and $S= A/4G\hbar$.) These physical considerations complemented the laws of event horizon mechanics. The close relation between black hole mechanics and thermodynamics is one of the most intriguing revelations to emerge from strong gravity. Indeed, even after five decades, its deeper meaning continues to be investigated.

\vskip0.3cm
\textbf{3. Beyond Event Horizons} 
\vskip0.2cm 
For almost three decades after the seminal results of Penrose and Hawking, event horizons were synonymous with black hole boundaries. Much of the discussion (and confusion) on the issue of `information loss' continues to be centered on the identification of black hole entropy with the area of cross sections of event horizons. But over the last quarter century it has become increasingly clear, through the work of many researchers, that this identification is deeply flawed for dynamical black holes (see, e.g., \cite{hayward2000blackholesnewhorizons,Ashtekar:2003hk,Ashtekar:2004cn,Visser_2014}). 

Various limitations of event horizons have been discussed in the literature (for recent summaries, see, e.g., \cite{Ashtekar:2025LivRev,aps-grg}). The most important among these is that they are \emph{teleological}. This does not mean that they have a purpose or predict the future. Rather, it means that whether or not a space-time point lies inside the horizon can be decided only after one knows the \emph{complete future} of the gravitational system under consideration. As discussed above, an event horizon is the boundary between the set of events from which future-directed causal signals can eventually reach $\scrip$ and the set from which no such signal \emph{ever} can. Because the word \emph{ever} is in the definition, no measurement made only here and now can locate the event horizon! For example, an event horizon may have \emph{formed}, and \emph{grown}, in your room during the last few minutes, \emph{in anticipation} of a gravitational collapse in our galaxy, a million years in the future! But you cannot be sure until you have the knowledge of the space-time geometry all the way to the \emph{infinite} future.
\vskip0.3cm

\noindent
\begin{minipage}{0.28\linewidth}
  \centering
    \includegraphics[width=\linewidth]{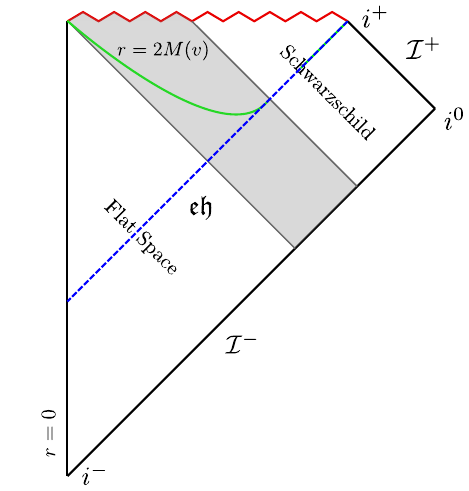}\vskip-0.4cm
    \hskip-0.2cm{\footnotesize{Fig. 2}} 
\label{fig2}
\end{minipage}
\hfill
\begin{minipage}{0.7\linewidth} 
\noindent  
Fig. 2 provides a concrete illustration of this spooky feature. It is the Penrose diagram of a Vaidya solution depicting again a gravitational collapse, but now of a null fluid that is incident from past infinity, $\scrim$. To the past of the infalling null fluid, the space-time is \emph{flat}. As one can see, the event horizon $\EH$ forms and grows already in this flat region,  although there is nothing happening there! It anticipates the collapse that is to occur in the future and creates a 
\emph{huge `black hole region' in a chunk of flat space-time}! A null fluid coming from $\scrim$ is not essential for this to occur. One can replace it with a suitable Vlasov fluid that comes from the past timelike infinity $i^{-}$ as in Fig. 1 \cite{KehleUnger2024}.
\end{minipage}
\vskip0.2cm

\noindent Since an event horizon can form and grow in flat regions of space-time in the absence of any physical process in the neighborhood of its growing segments, physically it is simply untenable to associate an entropic connotation with the growing area of its cross-sections! For thermodynamic considerations we need to replace event horizons with an alternative, more satisfactory notion, and find a better measure of entropy that is applicable to all dynamical black holes. Fortunately, this is possible: there \emph{are} more satisfactory characterizations of black hole boundaries, called \emph{Quasi-Local Horizons} (QLHs)  \cite{hayward2000blackholesnewhorizons,Ashtekar:2003hk}, that provide a viable notion of entropy even when black holes are far from equilibrium \cite{aps-prl,aps-grg}. Over the last two decades or so, QLHs have been used widely in mathematical relativity, numerical simulations of black hole mergers and quantum gravity (see recent reviews \cite{Ashtekar:2025LivRev,ashtekar2023regularblackholesloop,ashtekar2025blackholeevaporationloop} and references to original papers therein).

QLHs are defined using geometry in their immediate vicinity and make no reference to $\scrip$. Therefore they are \emph{free of teleology}. In the course of evolution, changes in fields on any of their segments occur only in response to physical processes happening on those segments themselves. In Fig. 2, the $r=2M(v)$ surface---depicted by a green line---denotes the segment of a QLH which forms and grows in area, directly because of the infalling matter, and then joins smoothly onto a segment of the event horizon in the future Schwarzschild region, once the infalling energy flux ends. This is a very simple illustration, but its qualitative features are generic. In particular, there are no QLHs in flat portions of \emph{any} space-time and the area of a given segment can grow (as in a black hole merger) or diminish (as during quantum evaporation) only because of fluxes of (positive or negative) energy across that segment. 

Because they are free of teleology, QLHs have had a broad range of applications. For example, during numerical simulations of black hole mergers, event horizons $\EH$ cannot be used to locate the progenitors, nor the remnant, since $\EH$ can only be found, as an afterthought, at the end of the simulation when one has the entire space-time. Therefore, waveform catalogs are created using QLHs and the numerical calculations of the masses and spins of the progenitors and the remnant are carried out on QLHs as well. Similarly, in the context of quantum gravity, as H\'aj\'i\v{c}ek \cite{PhysRevD.36.1065} pointed out in 1987, one can get rid of event horizons completely by modifying geometry in a Planck scale neighborhood of the singularity. In 2004, during the GR-17 conference, in the context of black hole evaporation Hawking himself emphasized that ``a true event horizon never forms". If the singularity is resolved by quantum gravity effects---as, e.g., in loop quantum gravity---then what forms during the gravitational collapse and evaporates due to quantum processes is a QLH \cite{ashtekar2025blackholeevaporationloop}.

Now, starting in the mid-1970s, event horizons have had dominant status in the theoretical physics community, primarily because of their thermodynamic connotations. (Indeed, much of the discussion on the issue of information loss in the high energy physics community has continued to be deeply rooted in event horizons.) So what happens to thermodynamic considerations if event horizons are abandoned in dynamical situations? Interestingly, thermodynamic considerations extend naturally to QLHs. In fact, in dynamical situations they provide a generalized first law and a quantitative second law, both with direct physical bearing. Entropy is now associated with certain cross sections (marginally trapped surfaces) of QLHs. When the QLH is quiescent, the area is constant while it changes if there are fluxes of energy across any of its segments. In place of the qualitative statement (\ref{2law1}) of area increase on event horizons, one has a \emph{quantitative} relation between the area change and a specific `entropic flux' across any segment of a QLH. The precise expression of the flux receives contributions from both matter and gravitational waves, but just on that segment \cite{aps-prl,aps-grg}. 

Finally, the notion of QLHs also involves only the space-time geometry. Yet rich physics emerges from them, offering new insights into black-hole dynamics and illuminating unforeseen features. They provide yet another example of the marvelous ability of general relativity to encode subtle physics in  space-time geometry.
%
 
\section{Gravitational waves beyond perturbation theory}
\label{s3}

Thanks to Maxwell and Hertz, the existence of electromagnetic waves was well established in the late 1800s. From a theoretical perspective, they are freely propagating electromagnetic fields, represented by source-free solutions to Maxwell's equations. The two polarization modes represent the two degrees of freedom of the electromagnetic fields that lead to the two helicities of photons in the quantum theory. 

By contrast, in Newtonian physics the gravitational field vanishes in the absence of matter sources; it does not have degrees of freedom of its own. What about general relativity that `relativizes' gravity? Does the gravitational field now acquire degrees of freedom of its own? Do gravitational waves exist in Nature? In the light of the LIGO-Virgo discoveries over the past decade that have been at center stage of fundamental physics and astronomy, this may seem like a strange question. However, this very issue remained controversial for some five decades after the discovery of general relativity. As with black holes, this is a fascinating piece of history. 
\vskip0.3cm

\vskip0.3cm
\textbf{1. Existence of Gravitational Waves}
\vskip0.2cm

Soon after his discovery of general relativity, Einstein investigated this issue using the weak-field approximation around Minkowski space-time and showed that a time-dependent mass quadrupole moment creates gravitational waves that propagate with the speed of light \cite{Einstein1918}. Later, Eddington further analyzed these waves, distinguishing the physical degrees of freedom from gauge modes, and showed that gravitational waves have two polarizations, just as electromagnetic waves do \cite{Eddington2016}.
\footnote{There is a common myth that Eddington regarded (linearized) gravitational waves as being unphysical, propagating with the speed of thought. It is incorrect. He clearly understood that the physical modes propagate with the speed of light; `speed of thought' refers to gauge modes.}
But the question of whether \emph{full, nonlinear general relativity} admits gravitational waves continued to be debated for decades. Since space-time diffeomorphisms represent gauge in the full theory, it was not clear whether what seemed like `wave-like' features in a solution to Einstein's equations were not mere coordinate artifacts. A concrete example illustrating the subtleties is the source-free solution discovered by  Levi-Civita \cite{LeviCivita1918}---called the c-metric in the Ehlers-Kundt classification--- in which the metric appears to admit gravitational waves. On the other hand, the metric also seemed to be stationary because it appeared to be time-independent in a certain coordinate system. Since stationary solutions cannot admit physical waves, could the apparent wave-like behavior simply be a coordinate artifact? 

Interestingly, two decades later Einstein himself contributed to this controversy! A joint paper with Rosen, entitled \emph{``Do Gravitational Waves Exist?''}, submitted to Physical Review, suggested that they \emph{don't exist} in full general relativity. In a letter to Born on this work, he said:\vskip0.05cm
\textsl{``Together with a young collaborator, I arrived at the interesting result that gravitational waves do not exist, though they had been assumed a certainty to the first approximation. This shows that the nonlinear general relativistic wave field equations can tell us more or, rather, limit us more than we had believed up to now". }\vskip0.05cm
However, he then realized the flaw in the argument and reversed the conclusion in the final version that appeared in a rather obscure journal \cite{EinsteinRosen1937}, now entitled \emph{``On Gravitational Waves''}. Solutions found in that paper are now called the Einstein-Rosen cylindrical waves. Interestingly, Rosen continued to believe in their non-existence well into the 1970s. (See, e.g., \cite{Kennefick1999,Kennefick2005} for this fascinating piece of history.) 

In the broader community, the controversy about the physical reality of gravitational waves continued until the early 1960s. In the late 1950s, Bondi had the key insight that, for isolated gravitating systems in full general relativity, one should move away from the sources along retarded null cones. Then, one can avoid the trap of coordinate ambiguities and isolate the physical gravitational waves from the $1/r$-departures of the solution from the flat metric. This idea was implemented in full generality by Sachs, Penrose, Newman and others and, by the late 1960s, it came to be widely accepted that gravitational waves do exist in full general relativity; they are not an artifact of the linearization approximation that Einstein had used in 1918. 

In recent years, their properties have been analyzed in detail using future null infinity $\scrip$, introduced in Section \ref{s2}. It serves as the natural arena to unambiguously disentangle coordinate artifacts from physical effects and write down invariant expressions, e.g., for the energy-momentum carried by gravitational waves.  
\footnote{In particular, one can revisit the Levi-Civita c-metric using this framework. It was shown that the metric does admit a $\scrip$ and gravitational waves carry energy-momentum across it \cite{KinnersleyWalker1970,AshtekarDray1981}. The apparent contradiction with `stationarity' is resolved by the fact that the symmetry in question represents a boost rather than a time translation. This analysis also showed that the c-metric represents the gravitational field of two black holes that are constantly accelerating but fail to merge into one another because there is a two-dimensional sheet representing a (unphysical) `spring' between them that cannot be compressed beyond a certain point.}
As in the linear theory, they have two true degrees of freedom encoded in two polarization modes. Recent discoveries by LIGO-Virgo collaborations have confirmed that they travel at the speed of light, as predicted by general relativity. This discovery has also served to rule out a number of alternative theories of relativistic gravity.
\vskip0.3cm

\textbf{2. Enlargement of the Symmetry Group and its Consequences}
\vskip0.2cm

Gravitational waves can be envisaged as ripples propagating on the very fabric of space-time. Because gravity is inseparably intertwined with geometry in general relativity, one faces new conceptual issues. They are so subtle that their consequences escaped even some of the leading figures ---including luminaries such as Feynman. The key reason is that while investigating non-gravitational interactions---such as electromagnetism--- one works with a fixed Minkowski metric $\mathring{\eta}_{ab}$ which is completely insensitive to the specific solutions to dynamical equations being considered. Using this `inert', background geometry, it is easy to introduce physically motivated boundary conditions, such as the $1/r$ fall-off of the dynamical fields of interest, and then arrive at well-defined expressions for the energy, momentum, and angular momentum carried by electromagnetic waves, as one moves away from the sources. In general relativity, by contrast, the physical metric plays a dual role: it determines space-time geometry and at the same time it encodes the gravitational field being investigated. Consequently, the meaning of `the $1/r$ fall-off' is tied to the curved metric which is itself the dynamical variable on which we wish to impose the boundary conditions. This nuance creates unfamiliar layers of complexity. 

In perturbative investigations of gravity, it is again natural to fix a Minkowski metric $\mathring{\eta}_{ab}$ as the background geometry and use it to define the $1/r$ fall-off of the first-order gravitational perturbation, and then add perturbative corrections order by order using the coupling constant $G$ of the theory. In this framework, the symmetry group of the theory to any given order is just the Poincar\'e group $\mathring{\mathcal{P}}$ of $\mathring{\eta}_{ab}$. Using the four-dimensional translation subgroup of $\mathring{\mathcal{P}}$ one can define the energy-momentum carried by gravitational waves at any given order. Similarly, using its (four-parameter family of) Lorentz subgroups, one can define the relativistic angular momentum of gravitational waves. In the quantum theory, the one-particle Hilbert space is constructed from linearized gravitational fields. The Poincar\'e group  $\mathring{\mathcal{P}}$ has a natural action on it. Using the Casimir operators of $\mathring{\mathcal{P}}$, one can compute the mass and spin and conclude that gravitons have mass zero and spin two. 

However, at the full, non-perturbative level this strategy faces a key conceptual problem. Given any gravitating isolated system ---such as a pair of black holes orbiting each other--- one only has the physical metric $g_{ab}$ which approaches flatness as one moves away from sources in null directions ---i.e., approaches $\scrip$. However, there is no invariant way to extract a \emph{preferred} Minkowski metric $\mathring{\eta}_{ab}$ from the physical $g_{ab}$! It \emph{is} true that since the curvature of $g_{ab}$ tends to zero as one recedes from sources, it approaches \emph{some} Minkowski metric. More precisely, asymptotic flatness implies that there exists a Minkowski metric $\eta_{ab}$ such that, in the $(t,r,\theta,\varphi)$ chart of that $\eta_{ab}$, the components of $(g_{ab} -\eta_{ab})$ fall off as $1/r$ as one recedes from sources and approaches $\scrip$. However, this $\eta_{ab}$ is \emph{not} unique! One can perform an \emph{angle-dependent} translation ---say $t \rightarrow t^\prime = t + f(\theta,\varphi)$, and $(r^\prime,\theta^\prime,\varphi^\prime) = (r,\theta,\varphi)$---and obtain \emph{another} Minkowski metric $\eta^\prime_{ab}$ using $(t^\prime, r^\prime,\theta^\prime,\varphi^\prime)$ (on the same space-time manifold). Then one can check that the components of $(g_{ab} - \eta^\prime_{ab})$ \emph{also fall off as} $1/r^\prime $. Thus, we will have two \emph{distinct} Poincar\'e groups ---$\mathcal{P}$ defined by $\eta_{ab}$ and $\mathcal{P}^\prime$ defined by $\eta^\prime_{ab}$---neither of which is preferred by the \emph{physical} metric $g_{ab}$. Consequently, the requirement that $g_{ab}$ is asymptotically flat does not provide us a single Poincar\'e group but a whole family of them, all of which preserve the asymptotic-flatness boundary conditions satisfied by $g_{ab}$. Therefore they \emph{all} represent asymptotic symmetries! 

How many are there? A detailed examination shows that there are as many as there are angle-dependent translations ---which are called supertranslations. This supertranslation freedom can be parametrized by smooth (but otherwise arbitrary) functions $f(\theta,\varphi)$ on a two-sphere. Roughly, the asymptotic symmetry group is obtained by putting together all these Poincar\'e groups in a coherent manner. It is called the Bondi-Metzner-Sachs (BMS) group, denoted by $\mathcal{B}$. It is infinite-dimensional. It has the same structure as the Poincar\'e group, with the four-dimensional (Abelian) subgroup of translations being replaced by the infinite-dimensional (also Abelian) subgroup of supertranslations. For asymptotically flat space-times in which there are no gravitational waves ---in particular in stationary space-times--- one can get rid of the supertranslation enlargement and find a canonical Poincar\'e subgroup $\mathcal{P}$ of the $\mathcal{B}$ using just the asymptotic structure. Thus, the tremendous enlargement of the (asymptotic) symmetry group from the ten-dimensional $\mathring{\mathcal{P}}$ of Minkowskian field theories to the infinite-dimensional $\B$ is a direct consequence of the fact that gravitational waves in full general relativity leave certain ripples on the space-time geometry that persist all the way to $\scrip$. As a consequence, the physical metric  asymptotically approaches a \emph{family of} Minkowski metrics (related by supertranslations) that differ from each other even to the leading asymptotic order. (For a review, see, e.g., \cite{aa-yau} and references therein.)

Interestingly, even in the presence of gravitational waves, translations can be singled out invariantly in the supertranslation subgroup. Therefore, one can \emph{unambiguously} define the energy-momentum carried away by gravitational waves. However, we also have infinitely many new observables, each associated with a genuine supertranslation. This richness has a direct physical implication that goes under the name of \emph{gravitational memory}. It is a measure of the permanent angle-dependent displacement that a detector undergoes after a generic gravitational wave passes through it. Interestingly, this effect has been observed in LIGO events, but so far its value is only `inferred' from observations rather than measured directly \cite{KheraKrishnanAshtekarDeLorenzo2021}. For a direct measurement we will likely have to wait for observations from the LISA space mission. Finally, while $\mathcal{P}$ admits a four-parameter family of Lorentz subgroups (`as many as there are translations'), $\B$ admits an infinite-parameter family (`as many as there are translations'). As a consequence, the notion of angular momentum carried by gravitational waves---or indeed, that carried by electromagnetic waves in a space-time that also admits gravitational waves---becomes much more involved \cite{PhysRevD.101.044005,AshtekarKhera2024}! These subtleties have been analyzed for LIGO events \cite{KheraAshtekarKrishnan2021}. 

To summarize, in the presence of gravitational waves the asymptotic symmetry group is no longer the familiar Poincar\'e group although physical space-times are asymptotically Minkowskian! This is a startling consequence of full nonlinear general relativity which escapes perturbative treatments because they work with a fixed Minkowski space-time in the background. 
One might argue that it should not matter which Minkowski background one chooses because any two Minkowski metrics on a given space-time manifold are gauge related. But this idea is naive: In the asymptotically flat context, diffeomorphisms (or coordinate transformations) correspond to gauge only if they are asymptotically identity. Therefore two Minkowski background metrics that differ even at infinity --such as the ones related by a supertranslation-- are related by \emph{an asymptotic symmetry}, rather than a gauge transformation. The consequent enlargement of the asymptotic symmetry group from a ten-dimensional Poincar\'e group to the infinite-dimensional BMS group has physical consequences. Specifically, implications of this enlargement can be and have been tested using gravitational wave observations. 

Finally, in spite of this huge enlargement of the symmetry group, one can still calculate the mass and spin of elementary quanta of radiative modes of exact general relativity, using the full BMS group at $\scrip$, without having to resort to perturbation theory: the mass is zero and spin is two \cite{ashtekar1987asymptotic}! 
Once appropriate concepts are identified, general relativity astutely enables us to extract fundamental physics from rather abstract geometrical notions, without having to take recourse to the weak-field approximation.
\vskip0.3cm
\textbf{3. Gravitational Wave Tomography}
\vskip0.2cm

In this section, I will discuss a recent development (based on joint work with Neev Khera) that combines the quasi-local horizons discussed above with the gravitational waves now under consideration.  As mentioned already in the abstract, I will continue to skip technical caveats and attempt to present a broad picture, emphasizing essential features. 

A quasi-local horizon $\QLH$ can have dynamical segments, $\DHS$, across which there is influx of energy, as well as quiescent ones, $\IHS$, which are isolated. An elementary example is provided by the surface $r=2M(v)$ in Fig. 2 where its dynamical segment $\DHS$ lies in the region containing the null fluid. It is spacelike and expands as the collapse proceeds and more and more fluid falls into the black hole. Once the influx ends, $\DHS$ joins smoothly to the quiescent null portion $\IHS$, with constant area, which coincides with the final segment of the event horizon that lies in the Schwarzschild region. 

Similarly, in a black hole merger, the dynamical segment $\DHS$ of the remnant is spacelike and expands as gravitational waves fall into the black hole. It asymptotes to a null segment $\IHS$ in the distant future, which coincides with (the end-piece of) the event horizon of the Kerr black hole representing the final equilibrium state. The geometry of the entire QLH contains information about its shape and spin content, which is captured by two sets of invariantly defined numbers: shape multipoles, $\I_{\ell, m}$ and spin multipoles $\S_{\ell,m}$. On isolated portions $\IHS$ of the QLH, they are time-independent and can be evaluated on any of its two-sphere cross-sections $S$ (i.e., at any time). Dynamical segments $\DHS$ come with a natural slicing by marginally trapped surfaces $S$ \cite{Ashtekar:2005ez} and passage from one marginally trapped surface $S$ to another $S'$ can be regarded as `time evolution'. Multipole moments are now associated with each $S$ and denoted by $\I_{\ell, m}[S],\,\S_{\ell,m}[S]$. Thus, the multipole moments of $\DHS$ are `time-dependent', and the evolution of the shape and spin structure of $\DHS$ is captured by this dependence.%
\footnote{Incidentally, evolution along event horizons cannot be captured by such invariantly defined observables, first because event horizons fail to be even $C^1$ in physical situations of interest \cite{Giulini1998,Krolak1999}, and second because they don't admit a natural foliation.}
As the QLH becomes quiescent, the $\DHS$ approaches an $\IHS$ and the time-dependence of multipoles fades away. In particular, in a black hole merger, the $\DHS$ asymptotes to the Kerr $\IHS$ and its multipoles $\I_{\ell, m}[S],\,\S_{\ell,m}[S]$ approach $\I_{\ell, m}^{\rm kerr},\,\S_{\ell,m}^{\rm kerr}$.

Of particular interest is the so-called \emph{quasinormal mode} (QNM) \emph{regime} that emerges very soon after the merger in which the space-time metric is well approximated by a perturbed Kerr metric. The term `quasinormal' refers to the fact that the time dependence of these perturbations closely resembles \emph{damped} sinusoids (rather than being purely oscillatory as is the case for normal modes). This regime plays an important role in gravitational wave physics because the frequencies of these QNMs are completely determined by the geometry of the Kerr metric to which the physical metric asymptotes. Gravitational wave detectors provide us with waveforms they receive from compact binary mergers. To extract physics of the merger one creates templates of waveforms predicted by general relativity for various choices of initial conditions (using approximation methods and numerical simulations). By matching an observed waveform with these templates, one identifies the parameters that characterize the progenitors and the remnant, thereby providing details of the specific merger witnessed by the detectors.

All the concepts and observables used in this identification refer to the asymptotic regime of space-time, idealized by $\scrip$, which is \emph{very} far from the strong-field region around black holes themselves. Therefore, one would think that almost all the information about the detailed happenings in the strong-field region would be diluted during the long journey to the faraway asymptotic region, making it impossible to extract it just from the waveform at $\scrip$. Interestingly, as I now explain, this is not the case! 

First, associated with each supertranslation at $\scrip$, and a cross-section $C$ thereof, there is a `charge integral' $\Q_{\ell, m}[C]$ providing the value of the $(\ell,m)$-component of supermomentum. These charges are `proper' scalars, i.e., insensitive to the choice of orientation on space-time, just as the shape multipole moments $\I_{\ell,m}[S]$ of the horizon are. There are also `dual' supermomenta ${}^\star{\Q}_{\ell, m}[C]$ which are `pseudo-scalars', just as the spin multipoles $\S_{\ell, m}[S]$ are. These observables at $\scrip$ change in time in response to the outgoing flux of gravitational waves crossing $\scrip$, just as the horizon multipole moments change in time in response to gravitational waves falling into the black hole across $\DHS$. It turns out that (in the QNM regime) Einstein's equations provide us with a specific map between the time dependence of $(\Q_{\ell, m}[C],\, {}^\star{\Q}_{\ell, m}[C] )$ at $\scrip$ and the time dependence of $\I_{\ell,m}[S],\, \S_{\ell, m}[S]$! Finally, 
the time dependence of the charges at $\scrip$ can be easily computed using the theoretical waveform at $\scrip$ that best matches the observations. Putting these two facts together, one can obtain a complete picture of the \emph{evolution} of the shape and spin structure of the horizon from the waveform seen by the LIGO-Virgo detectors. Put differently, happenings recorded by the detector in the asymptotic region, far away from the black hole remnant, suffice to generate a complete description of how the horizon geometry evolves in space-time regions where curvature is strongest.

What is most striking about this finding is that \emph{no causal signal} can be sent to $\scrip$ from the QLH $\QLH$ (or vice versa). Nonetheless, in the QNM regime one can accurately image horizon dynamics knowing just the gravitational waveform at $\scrip$! Recall that in medical tomography one monitors changes of a tumor by using images created by scattering of electromagnetic or sound waves, without having to `see' the tumor directly through surgery. In the same vein, in general relativity one can create a movie of horizon dynamics, although one cannot `look at the horizon', now \emph{even in principle}. This gravitational wave tomography is made possible, again, thanks to Einstein's equations. For more detailed discussions, see \cite{aa-banff,Ashtekar:2025LivRev}.

\section {Discussion}
\label{s4}

In the last two sections I collected a few of the startling consequences of Einstein's equations. They reveal the subtlety of general relativistic effects in the strong-field regime that cannot be captured by perturbative treatments. I purposely focused on black holes and gravitational waves as these areas feature prominently in fundamental physics as well as in astronomy and astrophysics. But there are also some notable surprises in foundational issues of primary interest to other communities. I will conclude with two examples.

The first concerns the initial value formulation. An important task for physics is to predict the future, knowing the initial conditions. For example, in Newtonian gravity, if one knows the initial data at an instant of time, e.g., the initial position and velocity of a test particle in the gravitational field of a large body, one can determine its dynamical trajectory for all times. For the electromagnetic field, the initial-value framework is interesting even in the absence of sources: Given the electric and magnetic field at an instant of time,  Maxwell's equations enable us to predict their values at \emph{all} later times. This is manifest in Maxwell's original formulation, which provides explicit equations for evolution of the electric and magnetic fields. Einstein's vacuum equations, $R_{ab} \,-\, \frac{1}{2}\, R g_{ab} =0$, on the other hand, were written down using  the four-dimensional  metric $g_{ab}$ without any reference to `initial data' or its `time-evolution'. Indeed, for decades after their discovery, it was unclear whether they even admit a sensible initial value formulation! It was only in the 1950s that it became clear, largely due to sustained investigations by Yvonne Choquet-Bruhat, that they do! Moreover, the initial value problem is \emph{much more} subtle than in Minkowskian field theories. For, the notions of an `instant of time', light cones, and causal propagation are not known a priori; they are determined by the solution that \emph{results} from the evolution! Thus, Einstein's equations do enable us to predict the future knowing the initial conditions, as in any useful physical theory, even though this is far from being obvious from their form. The combination of underlying conceptual subtleties, `tightness' of the framework, and the deep underlying conceptual coherence that makes this possible is truly remarkable.

The second feature concerns the positivity of energy. In Newtonian gravity, the gravitational potential energy $V$ is negative. But special relativity tells us that a rest-mass $m$ of the gravitating particle also contributes to energy via  $E = mc^2$, which is positive (and `huge' because of the $c^2$ factor). Nonetheless, can we envisage a gravitational system with, say, two equal mass particles for which the rest mass energy $2mc^2$ is overshadowed by the negative potential energy so that the `total energy' $E = 2mc^2 + V$ is negative? A simple calculation shows that we can indeed achieve this by putting the particles close enough so that they are inside a ball of the (`Schwarzschild') radius $r = \frac{4G\, m}{c^2}$. In fact in Newtonian gravity one can find configurations in which the `total energy' is arbitrarily negative. 

But the appearance of the Schwarzschild radius immediately suggests that in this regime Newtonian gravity is inadequate since general relativistic `corrections' dominate. So the question arises: Is the total energy of an isolated gravitational system necessarily positive in general relativity? If it allowed configurations in which the total energy became arbitrarily negative, they would open a pathway to induce instability in most, if not all, gravitating systems. There could be generic `run-away' solutions in which the system evolves to states with arbitrarily large amounts of negative energy by continuously radiating away positive energy. So the question arises: Can we arrange ordinary, classical matter whose own energy is non-negative (i.e., its stress tensor satisfies an appropriate energy condition) but for which the total energy, including the negative gravitational potential energy, becomes negative? If this were possible, the theory would have a hard time describing our universe.

The answer to this question proved to be surprisingly difficult, in part because the notion of `total energy of an isolated system' is rather involved in general relativity (and indeed in all relativistic gravitational theories). The notion is necessarily global because the gravitational field itself gravitates and carries energy. As a consequence, it receives contributions from everywhere and one has to go all the way to spatial infinity to define it.%
\footnote{The precise expression of total energy was first introduced by Arnowitt, Deser and Misner \cite{ADM1962} in the early 1960s using a Hamiltonian framework. But it was not obvious as to why that specific quantity should have the \emph{physical} connotation of total energy. The \emph{operational} meaning of this rather mathematical definition was clarified using thought experiments only recently \cite{Ashtekar4_2024}.}
    There was no clear-cut consensus in the community and in the 1970s, leading mathematical relativists such as Robert Geroch and James York attempted to find counter-examples to what was then a `positive energy conjecture'. However, by the early 1980s, thanks to the results of Richard Schoen and Shing-Tung Yau \cite{SchoenYau1981}, and Edward Witten \cite{Witten:1981mf}, Thomas Parker and Clifford Taubes \cite{ParkerTaubes1982}, it was firmly established that the conjecture is in fact a theorem. It holds for isolated systems whose matter sources satisfy the appropriate energy condition, and thus also for space-times containing black holes. Again, an essential, physically desirable property of isolated gravitating systems is embedded in general relativity. But to tease it out from its deep structures, one has to develop sufficient intuition for the way the theory encodes fundamental physics in geometry!

I hope that these examples  of the alluring features of general relativity will entice younger researchers to explore yet other aspects of the the strong field regime of the theory and uncover further surprises.

\bigskip\bigskip
\textbf{Acknowledgments:} Over the years, I have benefited from discussions with a large number of colleagues on the material presented here. I would like to thank especially Robert Geroch, Neev Khera, Badri Krishnan and Roger Penrose. This work was supported in part by the Eberly and Atherton Funds of the Eberly College of Science.
\bibliography{JVN}

\end{document}